\documentclass[twocolumn,amsmath,showkeys,prb]{revtex4-1}

\usepackage{dcolumn}
\usepackage{bm}
\usepackage[T1]{fontenc}
\usepackage{amsmath}
\usepackage{amsfonts}
\usepackage{amssymb}
\usepackage{graphicx}
\usepackage{tabularx}

\usepackage{hyperref}
\hypersetup{colorlinks=true, linkcolor=blue, citecolor=blue, urlcolor=blue, pdftitle={Geometric Ferroelectricity in Fluoro-Perovskites}, pdfauthor={Eric Bousquet}}

\begin{document}
\title{Geometric Ferroelectricity in Fluoro-Perovskites}
\author{A. C. Garcia-Castro$^{1,2}$, N. A. Spaldin$^3$ A. H. Romero$^{4}$ and E. Bousquet$^{1}$}
\affiliation{$^1$Physique Th\'eorique des Mat\'eriaux, Universit\'e de Li\`ege, B-4000 Sart-Tilman, Belgium}
\affiliation{$^2$Centro de Investigaci\'on y Estudios Avanzados del IPN, MX-76230, Quer\'etaro, M\'exico}
\affiliation{$^3$Department of Materials, ETH Zurich, Wolfgang-Pauli-Strasse 27, CH-8093 Zurich, Switzerland}
\affiliation{$^4$Physics Department, West Virginia University, WV-26506-6315, Morgantown, USA}

\begin{abstract}
We used first-principles calculations to investigate the existence and origin of the ferroelectric instability in the \emph{AB}F$_3$ fluoro-perovskites.
We find that many fluoro-perovskites have a ferroelectric instability in their high symmetry cubic structure, which is of similar amplitude to
that commonly found in oxide perovskites. 
In contrast to the oxides, however, the fluorides have nominal Born effective charges, indicating a different
mechanism for the instability. 
We show that the instability originates from ionic size effects, and is therefore in most cases largely insensitive
to pressure and strain, again in contrast to the oxide perovskites. An exception is NaMnF$_3$, where coherent epitaxial strain matching to a substrate
with equal $in$-plane lattice constants destabilizes the bulk \emph{Pnma} structure leading to a ferroelectric, and indeed multiferroic, ground state 
with an unusual polarization/strain response.
\end{abstract}

\maketitle

Since its discovery in BaTiO$_3$, ferroelectricity in perovskite-structure oxides has attracted tremendous interest, ranging from fundamental studies 
to technological applications \cite{lines}. Indeed, the transition-metal/oxygen bond, with its large polarizability, is particularly 
favourable for promoting the transition-metal off-centering that can result in a ferroelectric ground state \cite{cohen1998, hill2000}. 
Ferroelectrics also exist, of course, in many material chemistries that do not contain oxygen, with a particularly extensive range of 
fluorine-based examples, including both polymers \cite{JLovinger1983}
and ceramics in many crystal classes (for a review see Ref. \onlinecite{scott2011}). Perhaps not surprisingly given the low polarizability of 
bonds with fluorine, the mechanisms for ferroelectricity in fluorine-based ferroelectrics are distinct from those in oxides, ranging 
from molecular reorientation in polymers \cite{Guo2013}, to geometric reconstructions in ceramics 
\cite{ederer2006}. These alternative mechanisms are of particular interest because, again unlike the oxides, they are
not contra-indicated by transition metal $d$ electrons, and so allow simultaneous ferroelectricity and magnetic ordering (multiferroism). 
Interestingly, however, none of the known perovskite-structure fluorides is reported to be ferroelectric. 

In this letter we use first-principles density functional calculations to investigate computationally the tendency towards ferroelectricity 
in perovskite-structure fluorides. Our goals are two-fold: First, by comparing to the behaviour of ferroelectric perovskite-structure
oxides, we further understanding of the driving forces for, and competition between, polar and non-polar structural distortions in 
both systems. Second, we aim to identify conditions, for example of strain or chemistry, under which ferroelectricity, and perhaps 
multiferroism, could be stabilized in the fluoride perovskites. 

\begin{figure}[htb!]
 \centering
 \includegraphics[width=8.5cm,keepaspectratio=true]{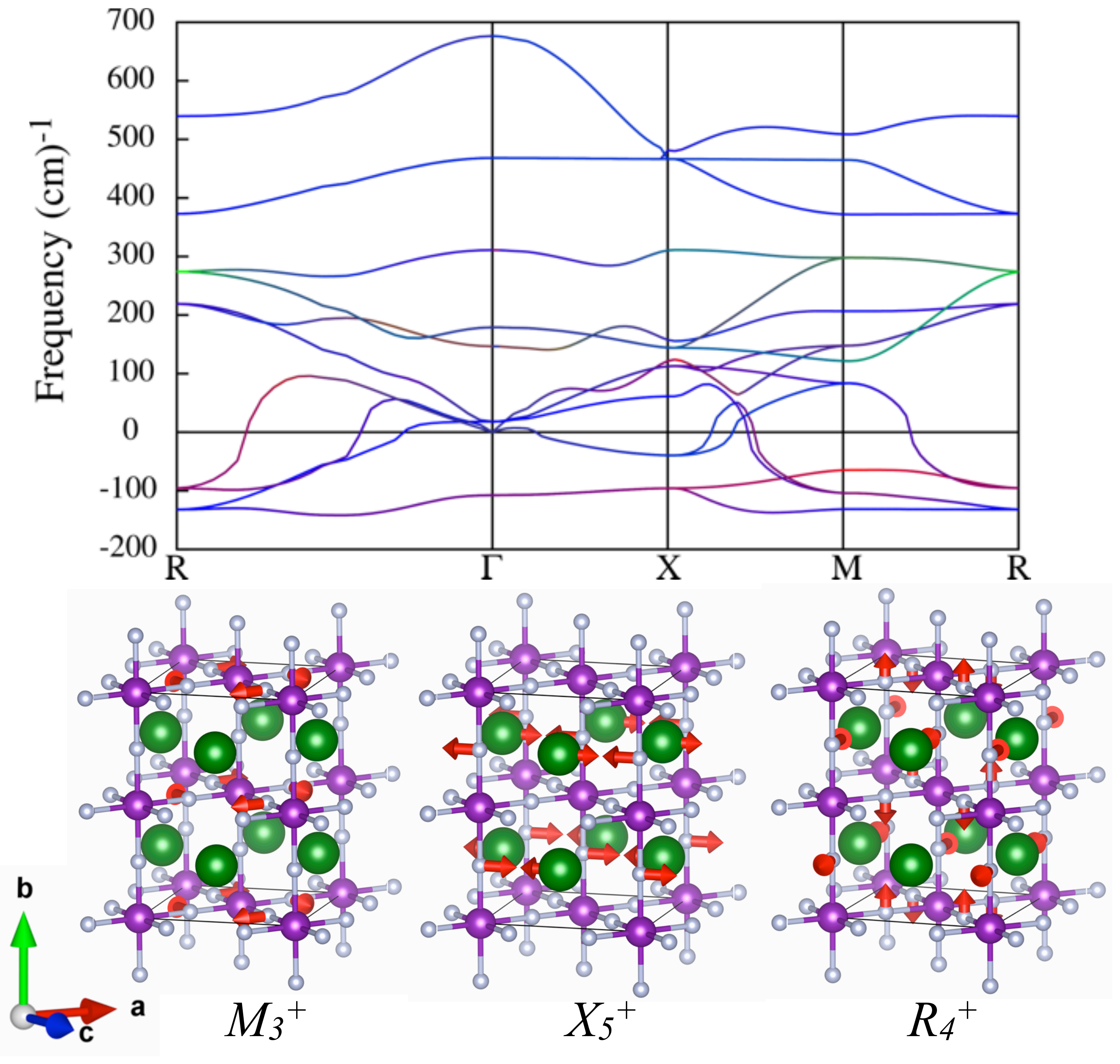}
 \caption{(Color online) Calculated phonon dispersion curves of cubic NaMnF$_3$ with a $G$-type AFM order. 
 The imaginary frequencies (unstable modes) are depicted as negative numbers.
 The branch colors are assigned according to the contribution of each atom to the dynamical matrix eigenvector: red for  Na, green for Mn, and blue for F. In the botton panel, the lower eigendisplacements at $M$, $X$ and $R$ for the modes that contribute most to the fully relaxed ground state are shown.}
 \label{fig:NaMnF3disp}
\end{figure}

A systematic study of the structural instabilities in the high symmetry cubic halide perovskites was carried out in the 1980's
using interionic potentials\cite{flocken1985}. Phonon frequencies were calculated at the $\Gamma$, $X$, $M$ and $R$ high symmetry
points of the cubic Brillouin zone for $A$$B$$X$$_3$, with $A$ = Li, Na, K, Rb or Cs, $B$ = Be, Mg or Ca, and $X$ a halide (F, Cl, Br or I).
While none of the perovskites has a ferroelectric ground state (although some, mostly Li, members of the series have the 
{\it non-perovskite} ferroelectric LiNbO$_3$ structure as their ground state), 24 of the 60 cases were found to show a ferroelectric (FE) 
instability. The instability was usually absent in compounds of the larger $A$ cations (which anyway usually form in a non-perovskite
hexagonal structure) and so it was concluded that the Na compounds should be the most promising candidates for ferroelectricity.
It was found, however, that in all cases a competing antiferrodistortive (AFD) zone boundary instability corresponding to rotations 
and tilts of the $X$ octahedra around the $B$ cation, dominated over the weaker polar instability, resulting in a centrosymmetric AFD 
ground state \cite{zhong1995}. Such competition between AFD and FE instabilities has been widely discussed in the perovskite
oxides literature \cite{vanderbilt1998, zhong1995, Benedek2013, Amisi2012} and recently shown, in 
the case of \textit{Pnma} perovskites to result from a mutual coupling to an $X_5^+$ mode involving antipolar displacements
of the $A$-cations and $X$ sites along the [101] axis as shown in  Fig. \ref{fig:NaMnF3disp}. 
Indeed artificially removing the $X_5^+$ mode in the calculations for perovskite oxides was shown to induce a ferroelectric ground state\cite{Benedek2013}. 

Our density functional calculations of the electronic and structural properties were performed using the PAW VASP code \cite{vasp1}, 
with the GGA PBEsol exchange correlation functional \cite{Perdew2008}, an energy cutoff of 700 eV on the plane wave expansion and a Brillouin-zone k-point 
sampling of 8$\times$8$\times$8 within the five 5-atom unit cell. Born effective charges and phonons were calculated using density 
functional perturbation theory \cite{gonze1997} in a 2$\times$2$\times$2 supercell. The dynamical matrix was unfolded using the 
Phonopy code\cite{phonopy} and the phonon dispersion curves were interpolated from the interatomic force constants (IFCs) using the 
$anaddb$ module provided with the Abinit software\cite{abinit}. 

In Fig. \ref{fig:NaMnF3disp} we show our calculated phonon dispersion for cubic NaMnF$_3$ with its ground state $G$-type 
antiferromagnetic (AFM) order. Unstable phonons are shown as negative frequencies. We find a polar instability
(of $\Gamma_4^-$ symmetry) at the zone center with eigendisplacement $\eta$ = [0.174, -0.003, -0.015, -0.094] for [Na, Mn, F$_\perp$, F$_\parallel$] 
respectively, and frequency 107$i$ cm$^{-1}$. The dispersion of this instability is small along $\Gamma-X-M$, similar to the behavior of typical perovskite oxide ferroelectrics \cite{ghosez1999}. In complete contrast to the oxides, however, this branch remains unstable at the $R$-point ($R_5^+$ mode)
where it corresponds to antiferroelectric (AFE) atomic displacements with each ``up'' dipole surrounded by neighboring ``down'' dipoles
in all three cartesian directions. 
Interestingly, this type of AFE instability has not been previously explored in perovskite systems \cite{KRabe2013a}.
The composition of the mode also differs from the perovskite oxides, as it is dominated by displacement of the
$A$-site cation, whereas in the oxides $B$-site displacements dominate. The eigendisplacement of the soft polar mode at $\Gamma$ in
BaTiO$_3$, for example, is $\eta$ = [0.001, 0.098, -0.071, -0.155] for [Ba, Ti, O$_\perp$, O$_\parallel$].
In addition to the unstable polar mode, we find strong AFD instabilities\cite{glazer1972} at the $M$ ($in$-phase rotation of the octahedra) and $R$ ($out$-of-phase rotation of the octahedra around) points, which have similar counterparts
in many perovskite oxides.

\begin{table}[htbp!]
\caption{Relaxed GGA PBEsol cell parameter, $a_0$ (\AA), $B$-radius (pm) and amplitude of the FE and AFD unstable modes 
(cm$^{-1}$) at the $\Gamma$, $X$, $M$ and $R$ points of selected Na$B$F$_3$ and $A$NiF$_3$ perovskites.
For  comparison we report the data for BaTiO$_3$.}
\centering
\begin{tabularx}{\columnwidth}{X c c c c c c c}
\hline
\hline 
system    & $a_0$ & $B$-radius & $\Gamma_4^-$ & $X_5^+$ & $R_5^+$ & $M_3^+$ & $R_4^+$    \rule[-1ex]{0pt}{3.5ex} \\
\hline
NaMnF$_3$ & 4.144 & 97         & 107$i$    & 96$i$ & 94$i$  & 132$i$    & 131$i$  \rule[-1ex]{0pt}{3.5ex}\\
NaVF$_3$  & 4.112 & 93          & 93$i$       & 95$i$ & 92$i$   &  131$i$   & 132$i$  \rule[-1ex]{0pt}{3.5ex}\\
NaZnF$_3$ & 4.000 & 88         & 85$i$      & 70$i$ &  59$i$  &   129$i$  & 130$i$  \rule[-1ex]{0pt}{3.5ex}\\
NaNiF$_3$ & 3.925 & 83          & 50$i$      & 47$i$ &  15  &   126$i$  & 127$i$  \rule[-1ex]{0pt}{3.5ex}\\
\emph{BaTiO$_3$} & \emph{3.955}    &      \emph{ 75}    &      \emph{ 220$i$}  & \emph{189$i$} & --- & \emph{ 165$i$} & ---        \rule[-1ex]{0pt}{3.5ex}\\
\hline
system    & $a_0$ & $A$-radius & $\Gamma_4^-$ & $X_5^+$ & $R_5^+$ & $M_3^+$ & $R_4^+$    \rule[-1ex]{0pt}{3.5ex} \\
\hline
CsNiF$_3$ & 4.155 & 181          & 128      & 226 &    258 &   193  & 184  \rule[-1ex]{0pt}{3.5ex}\\
RbNiF$_3$ & 4.063 & 166          & 133      & 212 &    258 &   153  & 149  \rule[-1ex]{0pt}{3.5ex}\\
KNiF$_3$ & 4.002 & 152          & 145      & 111 &    94 &   97  & 142  \rule[-1ex]{0pt}{3.5ex}\\
LiNiF$_3$ & 3.877 & 90          & 224$i$      & 213$i$ &  166$i$ &   190$i$  & 214$i$  \rule[-1ex]{0pt}{3.5ex}\\
\hline
\hline
\end{tabularx}
\label{tab:freq}
\end{table}

In Table~\ref{tab:freq} we list the frequencies of the unstable modes at the high symmetry points for the NaMnF$_3$ case
of Fig.~\ref{fig:NaMnF3disp} as well as for series obtained by first changing the $B$ site, then the $A$ site.
We note that the $R_4^+$ and $M_3^+$  instabilities (which are the AFD $out$-of-phase and $in$-phase rotations of the octahedra) are almost unaffected by the change of the $B$ site. 
However, the frequencies of the ferroelectric $\Gamma_4^-$, and the antiferroelectric $X_5^+$ and $R_5^+$ instabilities vary with the $B$ site, with larger radius $B$ cations having larger cell volumes and in turn stronger FE and AFE instabilities.  
Interestingly, we find quite different behaviour for the $A$ sites: Both $R_4^+$ and $M_3^+$ AFD modes, and 
$X_5^+$ and $R_5^+$ AFE modes show a strong dependence on the $A$-site, with smaller $A$-site cations having softer (or unstable) AFE and AFD phonons, in spite of the reduced cell volume. The volume dependence of the AFD modes follows the same trend as in perovskite oxides, in which
rotational instabilities are suppressed as the tolerance factor increases \cite{Benedek2013}. The 
FE $\Gamma_4^-$ mode shows no obvious trend with $A$-site radius, but is strongly unstable in LiNiF$_3$.
This behavior is distinct from the oxides, where the increase of volume associated with larger $A$ cations
tends to strengthen the FE instability \cite{ghosez1996}. 

\begin{figure}[htb!]
 \centering
 \includegraphics[width=8.0cm,keepaspectratio=true]{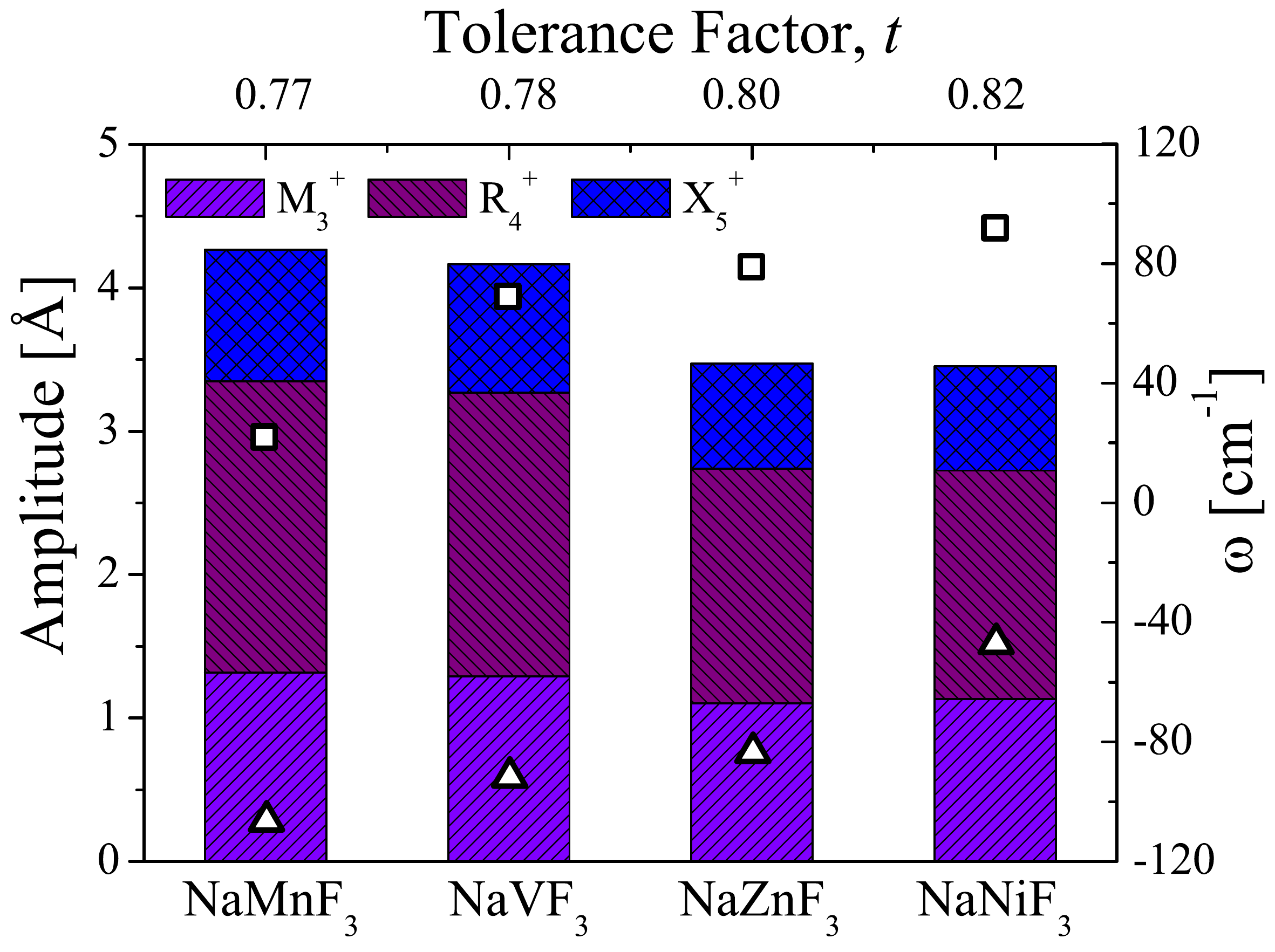}
 \caption{(Color online) Contributions of the unstable phonon modes of the cubic high symmetry structures to the fully relaxed ground states 
of \textit{Pnma} Na$B$F$_3$ fluorides.\cite{Orobengoa2009} Also shown are the frequency values for the FE modes in the \textit{Pnma} (white squares)
and cubic (white triangles) structures.} 
 \label{fig:modes-decomp}
\end{figure}

As a next step in the analysis, we decompose the structural distortions that map the high symmetry cubic perovskite structures 
into their fully relaxed \textit{Pnma} ground states into linear combinations of the unstable phonons of the
cubic structure \cite{Orobengoa2009}. 
Our results are shown in Fig. \ref{fig:modes-decomp} for the Na$B$F$_3$ series with $B$ = Mn, V, Zn and Ni.
As expected, and as is the case for \textit{Pnma} perovskite oxides, we find that the strongest contributions come from the two 
($R_4^+$ and $M_3^+$) unstable AFD modes. 
The absolute magnitudes of the contributions decrease across the series as the
$B$ site cation becomes smaller, consistent with the reduction in the amount of the tiltings and rotations when the tolerance factor increases. 
The next largest contribution is from the $X_5^+$ mode with its anti-polar $A$-site motions; notably its
contribution to the relaxed ground state is approximately double that of the same mode in the perovskite oxides.
Again, the magnitude of its contribution is smaller for smaller $B$ site cations. There are also small contributions from
the $R_5^+$ and $M_2^+$ modes which are roughly constant across the series and which we do not show.

We also show in Fig.~\ref{fig:modes-decomp} our calculated frequency for the unstable FE mode in the cubic structure (white triangles)
as well as the frequency of the FE soft mode in the $Pnma$ ground state (white squares). 
We see that the frequencies of the soft modes in both the cubic and the ground state structures scale with the tolerance factor: 
the smaller the tolerance factor, the smaller the FE mode frequency in the \emph{Pnma} structure, and the stronger the FE 
instability in the cubic phase.
We note also that a strong ferroelectric instability correlates with a large antipolar $X_5^+$ mode contribution, in complete
contrast to the case of perovskite oxides \cite{Benedek2013}.

To understand the differences that we have found between oxide and fluoride perovskites, we next calculate the Born effective charges ($Z^*$) of 
selected $A$$B$F$_3$ cubic perovskites, and compare them in Table~\ref{tab:bec} with the values for BaTiO$_3$. 
As expected for ionic compounds, the Born effective charges of the fluoro-perovskites are close to their nominal ionic charges,
with the most anomalous value ---a deviation of -0.71 \emph{e} for $F_\parallel$--- resulting from a small but non-zero charge transfer 
from F to $B$ as the atoms move closer to each other. 
This is in striking contrast to the oxides where the Born effective charges are strongly anomalous.
For example the corresponding oxygen anomaly in BaTiO$_3$ is -3.86  \emph{e} indicating
a much stronger dynamical charge transfer and change of hybridization accompanying the Ti-O displacement.
Since a key ingredient in explaining ferroelectricity in perovskite oxides is the presence of strongly anomalous Born effective 
charges on the $B$ and O atoms \cite{ghosez1998}, our results are consistent with the absence of ferroelectricity in the fluorides.
They do not explain, however, the sizeable FE instability that we find in $A$$B$F$_3$ {\it in spite} of the nominal Born effective charges.
We address this next.

\begin{table}[htbp!]
\caption{Born effective charges, $Z^*$ ($e$), and eigendisplacements of the FE unstable mode, $\eta$, in Na$B$F$_3$ and BaTiO$_3$.  
$X_\perp$ and $X_\parallel$ indicate the Born effective charge of the anion (F or O) when it is displaced parallel or perpendicular
to the $B-X$ bond.}
\centering
\begin{tabularx}{\columnwidth}{c c c c c | c c c c}
\hline
\hline 
  &                            &\multicolumn{2}{c}{$Z^*$}  & & &  \multicolumn{2}{c}{$\eta$}         \rule[-1ex]{0pt}{3.5ex} \\
system    & $A$  & $B$    & $X_\perp$ & $X_\parallel$ & $A$  & $B$    & $X_\perp$ & $X_\parallel$  \rule[-1ex]{0pt}{3.5ex} \\
\hline
Nominal   & 1    & 2      & -1      & -1      & --- & ---  & ---   & ---  \rule[-1ex]{0pt}{3.5ex}\\
NaMnF$_3$ & 1.17 & 2.21   & -0.83   & -1.72    & 0.174 & -0.005  & -0.020   & -0.088  \rule[-1ex]{0pt}{3.5ex}\\
NaVF$_3$  & 1.18 & 1.99   & -0.71   & -1.75     & 0.181 & -0.018  & -0.019   & -0.077 \rule[-1ex]{0pt}{3.5ex}\\
NaZnF$_3$ & 1.15 & 2.22   & -0.84   & -1.69     & 0.177 & -0.005  & -0.025   & -0.084 \rule[-1ex]{0pt}{3.5ex}\\
NaNiF$_3$ & 1.15 & 2.05   & -0.74   & -1.73     & 0.186 & -0.002  & -0.025   & -0.068 \rule[-1ex]{0pt}{3.5ex}\\
\emph{BaTiO$_3$} & \emph{2.75} & \emph{7.37}   & \emph{-2.14}   & \emph{-5.86}     & \emph{0.001} & \emph{0.098}  &\emph{ -0.071}   & \emph{-0.155} \rule[-1ex]{0pt}{3.5ex}\\
\hline
\hline
\end{tabularx}
\label{tab:bec}
\end{table}

\begin{figure}[htb!]
\centering
\includegraphics[width=8.5cm,keepaspectratio=true]{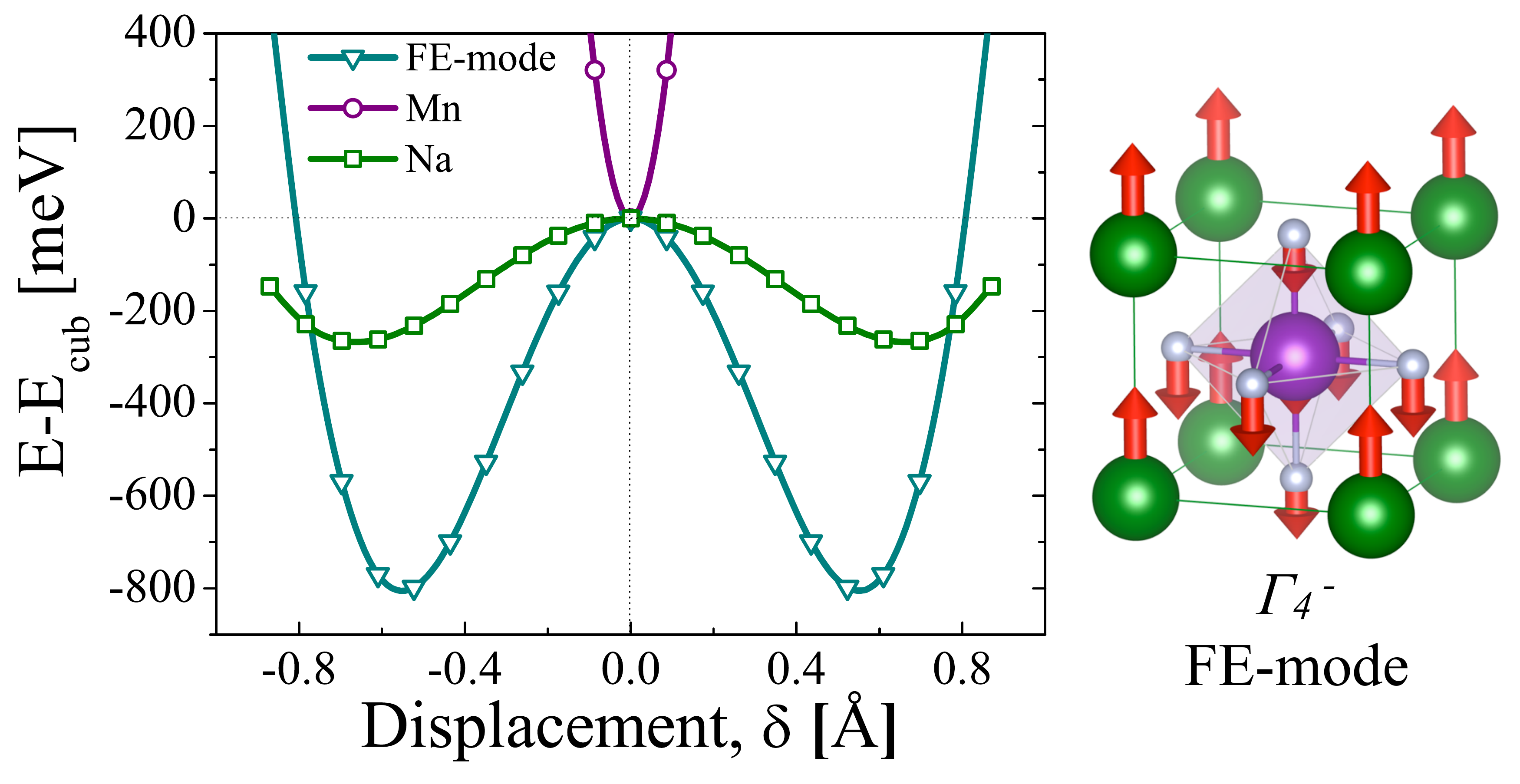}
\caption{(Color online) Left: Energy as a function of displacement for NaMnF$_3$ with a $G$-type antiferromagnetic order for three
different displacement patterns: The purple line corresponds to displacing only the Mn ions in each unit cell, the green to
displacing only the Na atoms, and the blue to the full eigendisplacement of the FE soft mode (right).} 
\label{fig:NaMnF3-double}
\end{figure}

We saw previously that the eigendisplacements of the ferroelectric instability in NaMnF$_3$ are strongly dominated by displacement of the Na ion. In Table \ref{tab:bec} we report the eigendisplacements\footnote{The eigendisplacements are normalized as follows: {$ \langle  \eta \lvert M \rvert \eta  \rangle = 1$}. $M$ is the atomic mass matrix $M_{ij} = \sqrt{M_i M_j}$ where $M_i$ is the mass of atom $i$.}  ($\eta$) of the FE unstable mode for the series of Na$B$F$_3$ compounds,
again comparing with BaTiO$_3$. In all fluoride cases we find a strong $A$-site contribution, with a substantial contribution
also from $F_\parallel$ but negligible contribution from the $B$-site ion. 
The dominance of the $A$-site displacement can also be seen in Fig. \ref{fig:NaMnF3-double}, where we show with red arrows the eigendisplacements for the FE-mode in NaMnF$_3$. The trend in $A$-site displacement magnitude reflects the $A$ and $B$-site relative masses: the $A$-site contributions to the FE mode eigendisplacements for $A$ = Li, Na, K, Rb and Cs in $A$NiF$_3$ are 0.347, 0.186, 0.138, 0.081 and 0.057 respectively while the 
contributions from Ni across the same series are 0.003, -0.019, -0.053, -0.073, -0.096.

To explore the origin of the unstable mode, we report in Table \ref{tab:IFC} the on-site interatomic force constants (IFC) for the 
cubic Na$B$F$_3$ series. The on-site IFC gives the force on an individual atom when it is displaced in the crystal while all the other atoms are kept fixed\footnote{The IFC matrix is defined by $C_{\alpha \beta i j}=\frac{\partial F_{\alpha i}}{\partial r_{\beta j}}$ where $i$ and $j$ refers to the directions, $\alpha$ and $\beta$ to the atoms, $F$ is the force on an atom and $r$ the atomic position. The acoustic sum rule imposes $\sum_\beta C_{\alpha \beta i i}=0$. The on-site IFC of atom $\alpha$ along direction $i$ corresponding to the matrix element $C_{\alpha \alpha i i}$ is thus equal to $C_{\alpha \alpha i i}=-\sum_{\beta\neq\alpha} C_{\alpha \beta i i}$ and so it can be seen as the resulting force on the atom $\kappa$ induced by all the other atoms when this atom $\kappa$ is displaced.}.
Even in high symmetry phases with structural instabilities, the on-site IFC are usually positive (indicating a restoring force) as
a periodic crystal is stable under the displacement of only one atom due to the breaking of the translational invariance.
This is indeed the case in all of the Na$B$F$_3$ compounds studied, however we note that the 
Na and F$_\perp$ on-site IFCs are unusually small. This indicates a small energy cost to displace an individual Na or F atom
away from its high symmetry position and is consistent with the large contribution of these atoms to the unstable eigendisplacements. 
We find that the magnitude of the on-site Na IFC depends on the size of the $B$ cation, with the smallest Na on-site IFCs found in
perovskites with larger $B$ cations and hence larger cell volumes and a roomier coordination cage around the Na atom. 
This size effect is confirmed by substituting other alkali atoms on the $A$ site. 
The progressively larger K, Rb and Cs atoms give on-site IFCs of 
2.88 eV\AA$^{-2}$, 4.16 eV\AA$^{-2}$ and 5.34 eV\AA$^{-2}$ respectively, while
in LiNiF$_3$, with its smaller Li $A$ site ion, the on-site IFC is negative (-0.34 eV\AA$^{-2}$).
The $A$-site on-site IFCs then follow the same volume trends as we observed previously for the
FE mode frequencies. 

\begin{table}[htbp!]
\caption{On-site IFCs (first four columns) and largest interatomic IFCs (last three columns) of $AB$F$_3$ (eV/\AA$^2$). $X_\perp$ and $X_\parallel$ are defined as in Table \ref{tab:bec}.}
\centering
\begin{tabularx}{\columnwidth}{X c c c  c | c c c}
\hline
\hline 
system    & $A$   & $B$    & $X_\perp$ & $X_\parallel$  &$B-X_\parallel$ & $B-B'_\parallel$ & $A-A'_\parallel$ \rule[-1ex]{0pt}{3.5ex} \\
\hline
NaMnF$_3$ & 0.17 & 13.14  & 0.76   & 13.35     & -4.62 & -2.77 & -0.51 \rule[-1ex]{0pt}{3.5ex}\\
NaVF$_3$  & 0.23 & 19.00  & 0.88   & 17.29     & -6.78 & -1.96 & -0.52 \rule[-1ex]{0pt}{3.5ex}\\
NaZnF$_3$ & 0.43 & 12.63  & 1.22   & 13.81     & -4.39 & -3.21 & -0.56 \rule[-1ex]{0pt}{3.5ex}\\
NaNiF$_3$ & 0.60 & 17.07  & 1.63   & 16.13     & -5.69 & -2.62 & -0.59 \rule[-1ex]{0pt}{3.5ex}\\
\emph{BaTiO$_3$}& \emph{8.27} & \emph{13.09} & \emph{6.74} & \emph{11.70}  & \emph{2.71} & \emph{-13.83} & \emph{-2.23} \rule[-1ex]{0pt}{3.5ex}\\
\hline
\hline
\end{tabularx}
\label{tab:IFC}
\end{table}

While the on-site IFCs of Na in Na$B$F$_3$ are all positive, we find that displacing  
all periodically repeated Na ions in the same direction (while keeping the $B$ and $F$ atoms fixed) lowers the energy of the system.
In Fig.  \ref{fig:NaMnF3-double} (green squares) we show the calculated total energy as a function of the magnitude of this collective
Na displacement in NaMnF$_3$. The ferroelectric characteristic double well shape is clear, with an energy lowering
of $\sim$ 200 meV (in the 40 atoms cell) The difference in behaviour between this collective displacement and that of an individual Na ion is due to the 
negative $A-A_\parallel$ IFC that is sufficient to destabilize the very small Na on-site IFC.
We find the opposite situation for all ions other than Na in the fluorides (the example of Mn in NaMnF$_3$ is shown in Fig. \ref{fig:NaMnF3-double}, purple circles). We suggest, therefore, that the FE instability
is driven by an instability dominated by the $A$-site cations.
Note, however, that while collectively displacing the Na ions lowers the energy in NaMnF$_3$, freezing in the full FE soft mode (blue triangles in Fig. \ref{fig:NaMnF3-double}) leads to an energy lowering that is four times larger indicating that the FE instability is a collective motion. 

\begin{figure}[htb!]
 \centering
 \includegraphics[width=7.5cm,keepaspectratio=true]{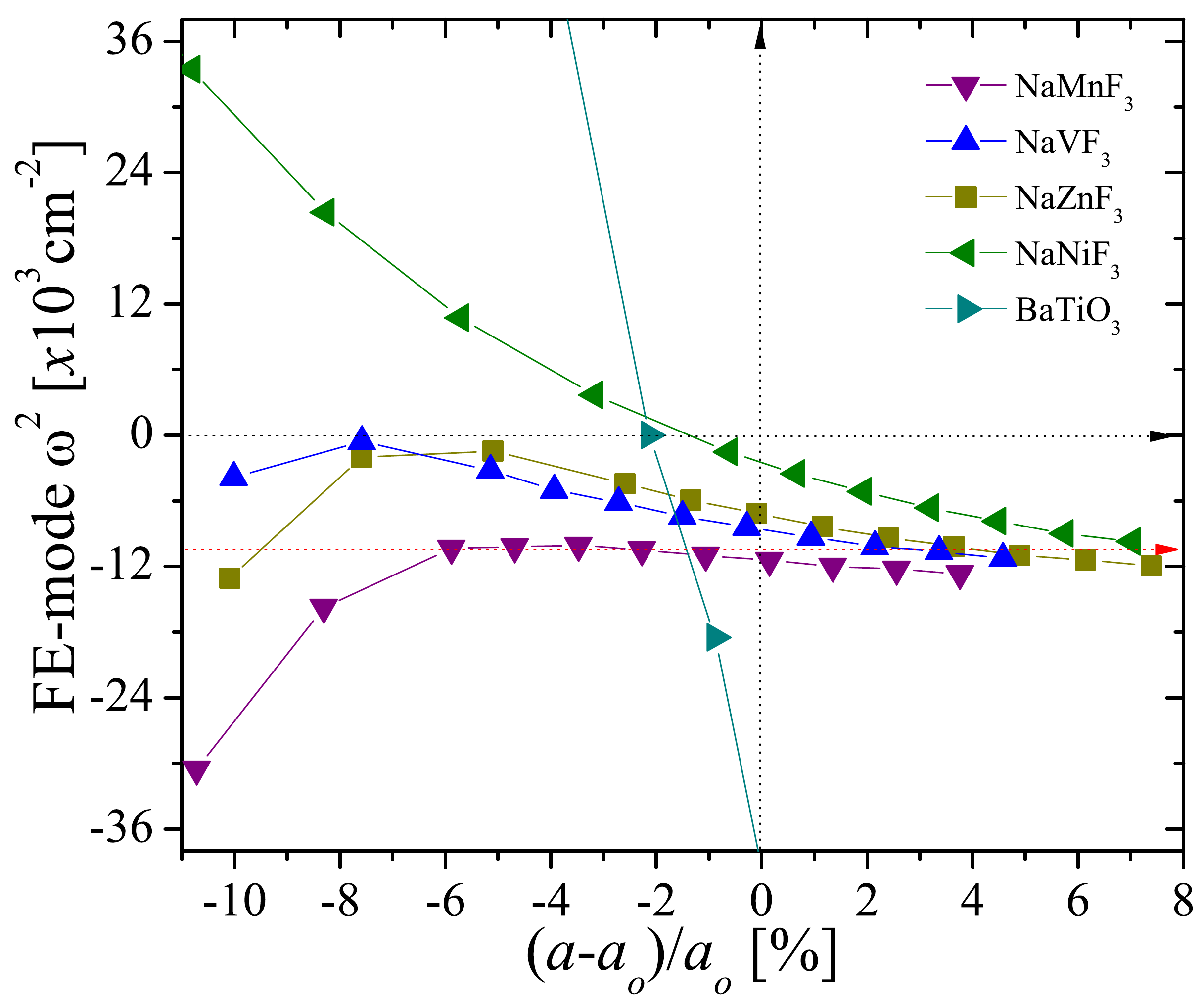}
  \caption{(Color online) Calculated frequency squared of the unstable FE phonon mode as a function of fractional change in cell parameter
for Na$B$F$_3$ with $B$ = Mn, V, Ni and Zn and BaTiO$_3$. $a$ is the varied cell parameter and $a_0$ is the PBEsol relaxed cell value.}
 \label{fig:omegavsP}
\end{figure}

Pursuing our comparison between oxides and fluoride perovskites, we next analyze the dependence of the FE instability on hydrostatic pressure.
In Fig. \ref{fig:omegavsP} we show our calculated square of the FE mode frequency ($\omega^2$) versus the cubic lattice parameter for Na$B$F$_3$ 
with $B$ = Mn, V, Zn and Ni as well as for BaTiO$_3$ \cite{kornev2007}.
We see immediately that the FE instability in the fluorides is much less sensitive to volume than in oxides.
For example, in BaTiO$_3$ a change of cell parameter of -2\% shifts $\omega^2$ by 40000 cm$^{-2}$ while the same change for NaMnF$_3$ 
gives a shift of only 2000 cm$^{-2}$. Second, the response of the fluorides is highly non-linear, and in some cases not even monotonic. And finally, for all compounds except NaNiF$_4$, compression enhances the FE instability, in stark contrast to the behavior in oxides.

\begin{figure}[htb!]
\centering
\includegraphics[width=8.0cm,keepaspectratio=true]{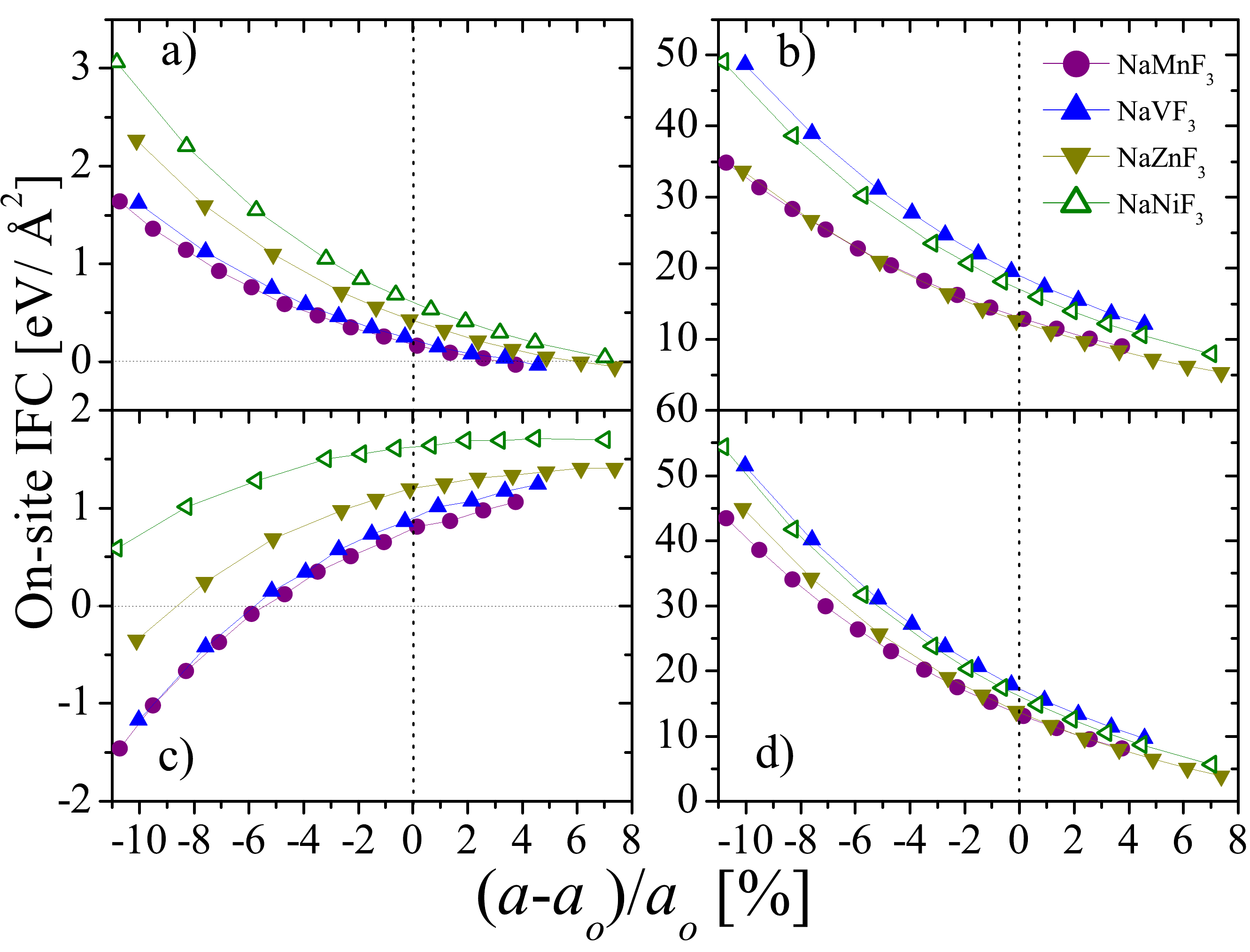}
\caption{(Color online) Evolution of the $A$ (panel a), $B$ (panel b), F$_\perp$ (panel c) and F$_\parallel$ (panel d) on-site IFCs (eV/\AA$^2$) with cubic cell parameter.}
\label{fig:IFCvsP}
\end{figure}

In Fig. \ref{fig:IFCvsP} we report the evolution of the four possible on-site IFCs of cubic Na$B$F$_3$ ($B$ = Mn, V, Zn and Ni) versus the pressure in order to explain this unusual pressure enhancement of the FE instability.
As expected from a simple ionic picture, the $A$ on-site IFC is stabilized under compression and further softened or destabilized on increasing the volume. 
The same trend is observed for the $B$ and F$_\parallel$ on-site IFCs while the F$_\perp$ on-site IFCs show the opposite behaviour.
Having a negative on-site IFC means there is no restoring force on the atom, thus the $A$$B$F$_3$ cubic structure is unstable to transverse on-site fluorine movements. This softening of the F$_\perp$ IFC is transferred into the FE mode by changing the character of its eigendisplacement, which evolves from $A$-site dominated toward F$_\perp$ dominated under pressure.
We note again a striking difference with BaTiO$_3$ , were none of the calculated on-site IFCs become unstable  even at extremely high pressure where ferroelectricity has been found to reappear\cite{kornev2007}.

Finally, we investigate whether it is possible to exploit this unusual ferroelectric instability to induce multiferroism in $A$$B$F$_3$.
It is well established that epitaxial strain can induce ferroelectricity in non-ferroelectric perovskite oxides of 
$Pnma$ and other symmetries \cite{gunter2012, Eklund2009}.
Therefore we explore whether similar strain-induced ferroelectricity can occur in $Pnma$ Na$B$F$_3$ taking the $ac$-plane as a plane for epitaxial strain.
We assume a cubic substrate that forces the $a$ and $c$ lattice constants to be equal, and
we relax the out-of-plane lattice parameter (orthorhombic $b$-axis) and internal atomic positions for each value of the in-plane
lattice constant. Note that the 0\% strain case corresponds to $a_{substrate}=\frac{a+c}{2}$, where $a$ 
and $c$ are the relaxed lattice parameters of the bulk $Pnma$ phase.
For all cases we find that the lowest FE mode has B$_{2u}$ symmetry (with frequencies 22 ,69, 79 and 92 cm$^{-1}$  for $B$ = Mn, V, Zn and Ni respectively, see Fig.  \ref{fig:modes-decomp}.) and is polarised perpendicular to the biaxial constraint.
We find that for all of them, this soft $B_{2u}$ mode is slightly sensitive to the value of strain, consistent with our earlier findings under pressure. However, we find that only for $B$ = Mn, the B$_{2u}$ is sufficiently low in frequency  to be destabilized by the epitaxial strain, and thus even at zero strain.  Performing structural relaxations yields a ferroelectric ground state (space group $Pna2_1$) with a $G$-type AFM and a polarization of 6 $\mu$C.cm$^{-2}$ at 0\% strain. Interestingly, negative or positive strain both enhance the polarization up to values of 12 $\mu$C.cm$^{-2}$ at +5\% and  9 $\mu$C.cm$^{-2}$ at -5\% strain, again in contrast with the behaviour of perovskites oxides.
This also means that the ground state of coherent epitaxially constrained NaMnF$_3$ is always multiferroic.
We note that the polarization develops perpendicular to the antipolar motions of the $X_5^+$ mode, in 
which the $A$ cations move in opposite directions in the $ac$-plane. 

In conclusion, we have investigated from first principles the origin of the FE instability in fluoro-perovskites.
We found that the FE instability in these ionic systems originates from the softness of the $A$-site displacements 
which in turn is caused by a simple geometric ionic size effect \cite{ederer2006}.
Due to its geometric origin, the fluoride FE instability is rather insensitive to pressure or epitaxial strain 
and most fluorides remain non-ferroelectric at all reasonable strain values. 
An exception is $Pnma$ NaMnF$_3$, in which the FE mode is particularly soft, so that it becomes ferroelectric and 
indeed multiferroic through coherent heteroepitaxy even at zero strain.
We hope that our results will motivate experimentalists to revisit the FE behavior of perovskite-structure
fluorides. 

We thank Ph. Ghosez for fruitful discussions.
This work was supported by the ETH-Z\"{u}rich, CONACYT Mexico under project 152153 and FRS-FNRS Belgium (EB). The computational resources
provided by the TACC supercomputer center and ETH-Z\"{u}rich Brutus Cluster are recognized.

\bibliography{biblio}

\begin{thebibliography}{28}%
\makeatletter
\providecommand \@ifxundefined [1]{%
 \@ifx{#1\undefined}
}%
\providecommand \@ifnum [1]{%
 \ifnum #1\expandafter \@firstoftwo
 \else \expandafter \@secondoftwo
 \fi
}%
\providecommand \@ifx [1]{%
 \ifx #1\expandafter \@firstoftwo
 \else \expandafter \@secondoftwo
 \fi
}%
\providecommand \natexlab [1]{#1}%
\providecommand \enquote  [1]{``#1''}%
\providecommand \bibnamefont  [1]{#1}%
\providecommand \bibfnamefont [1]{#1}%
\providecommand \citenamefont [1]{#1}%
\providecommand \href@noop [0]{\@secondoftwo}%
\providecommand \href [0]{\begingroup \@sanitize@url \@href}%
\providecommand \@href[1]{\@@startlink{#1}\@@href}%
\providecommand \@@href[1]{\endgroup#1\@@endlink}%
\providecommand \@sanitize@url [0]{\catcode `\\12\catcode `\$12\catcode
  `\&12\catcode `\#12\catcode `\^12\catcode `\_12\catcode `\%12\relax}%
\providecommand \@@startlink[1]{}%
\providecommand \@@endlink[0]{}%
\providecommand \url  [0]{\begingroup\@sanitize@url \@url }%
\providecommand \@url [1]{\endgroup\@href {#1}{\urlprefix }}%
\providecommand \urlprefix  [0]{URL }%
\providecommand \Eprint [0]{\href }%
\providecommand \doibase [0]{http://dx.doi.org/}%
\providecommand \selectlanguage [0]{\@gobble}%
\providecommand \bibinfo  [0]{\@secondoftwo}%
\providecommand \bibfield  [0]{\@secondoftwo}%
\providecommand \translation [1]{[#1]}%
\providecommand \BibitemOpen [0]{}%
\providecommand \bibitemStop [0]{}%
\providecommand \bibitemNoStop [0]{.\EOS\space}%
\providecommand \EOS [0]{\spacefactor3000\relax}%
\providecommand \BibitemShut  [1]{\csname bibitem#1\endcsname}%
\let\auto@bib@innerbib\@empty
\bibitem [{\citenamefont {Lines}\ and\ \citenamefont {Glass}(1977)}]{lines}%
  \BibitemOpen
  \bibfield  {author} {\bibinfo {author} {\bibfnamefont {M.~E.}\ \bibnamefont
  {Lines}}\ and\ \bibinfo {author} {\bibfnamefont {A.~M.}\ \bibnamefont
  {Glass}},\ }\href@noop {} {\emph {\bibinfo {title} {Principles and
  applications of ferroelectrics and related materials}}},\ \bibinfo {edition}
  {clarendon press}\ ed.\ (\bibinfo  {publisher} {Oxford},\ \bibinfo {year}
  {1977})\BibitemShut {NoStop}%
\bibitem [{\citenamefont {S{\'a}ghi-Szab{\'o}}\ \emph
  {et~al.}(1998)\citenamefont {S{\'a}ghi-Szab{\'o}}, \citenamefont {Cohen},\
  and\ \citenamefont {Krakauer}}]{cohen1998}%
  \BibitemOpen
  \bibfield  {author} {\bibinfo {author} {\bibfnamefont {G.}~\bibnamefont
  {S{\'a}ghi-Szab{\'o}}}, \bibinfo {author} {\bibfnamefont {R.~E.}\
  \bibnamefont {Cohen}}, \ and\ \bibinfo {author} {\bibfnamefont
  {H.}~\bibnamefont {Krakauer}},\ }\href {\doibase 10.1103/PhysRevLett.80.4321}
  {\bibfield  {journal} {\bibinfo  {journal} {Phys. Rev. Lett.}\ }\textbf
  {\bibinfo {volume} {80}},\ \bibinfo {pages} {4321} (\bibinfo {year}
  {1998})}\BibitemShut {NoStop}%
\bibitem [{\citenamefont {Hill}(2000)}]{hill2000}%
  \BibitemOpen
  \bibfield  {author} {\bibinfo {author} {\bibfnamefont {N.~A.}\ \bibnamefont
  {Hill}},\ }\href {\doibase 10.1021/jp000114x} {\bibfield  {journal} {\bibinfo
   {journal} {J. Phys. Chem. B}\ }\textbf {\bibinfo {volume} {104}},\ \bibinfo
  {pages} {6694} (\bibinfo {year} {2000})}\BibitemShut {NoStop}%
\bibitem [{\citenamefont {Lovinger}(1983)}]{JLovinger1983}%
  \BibitemOpen
  \bibfield  {author} {\bibinfo {author} {\bibfnamefont {A.~J.}\ \bibnamefont
  {Lovinger}},\ }\href {\doibase 10.1126/science.220.4602.1115} {\bibfield
  {journal} {\bibinfo  {journal} {Science}\ }\textbf {\bibinfo {volume}
  {220}},\ \bibinfo {pages} {1115} (\bibinfo {year} {1983})}\BibitemShut
  {NoStop}%
\bibitem [{\citenamefont {Scott}\ and\ \citenamefont
  {Blinc}(2011)}]{scott2011}%
  \BibitemOpen
  \bibfield  {author} {\bibinfo {author} {\bibfnamefont {J.~F.}\ \bibnamefont
  {Scott}}\ and\ \bibinfo {author} {\bibfnamefont {R.}~\bibnamefont {Blinc}},\
  }\href {\doibase 10.1088/0953-8984/23/11/113202} {\bibfield  {journal}
  {\bibinfo  {journal} {J. Phys. Condens. Matter}\ }\textbf {\bibinfo {volume}
  {23}},\ \bibinfo {pages} {113202} (\bibinfo {year} {2011})}\BibitemShut
  {NoStop}%
\bibitem [{\citenamefont {Guo}\ and\ \citenamefont {Setter}(2013)}]{Guo2013}%
  \BibitemOpen
  \bibfield  {author} {\bibinfo {author} {\bibfnamefont {D.}~\bibnamefont
  {Guo}}\ and\ \bibinfo {author} {\bibfnamefont {N.}~\bibnamefont {Setter}},\
  }\href {\doibase 10.1021/ma302377q} {\bibfield  {journal} {\bibinfo
  {journal} {Macromolecules}\ }\textbf {\bibinfo {volume} {46}},\ \bibinfo
  {pages} {1883} (\bibinfo {year} {2013})}\BibitemShut {NoStop}%
\bibitem [{\citenamefont {Ederer}\ and\ \citenamefont
  {Spaldin}(2006)}]{ederer2006}%
  \BibitemOpen
  \bibfield  {author} {\bibinfo {author} {\bibfnamefont {C.}~\bibnamefont
  {Ederer}}\ and\ \bibinfo {author} {\bibfnamefont {N.~A.}\ \bibnamefont
  {Spaldin}},\ }\href {\doibase 10.1103/PhysRevB.74.020401} {\bibfield
  {journal} {\bibinfo  {journal} {Phys. Rev. B}\ }\textbf {\bibinfo {volume}
  {74}},\ \bibinfo {pages} {020401} (\bibinfo {year} {2006})}\BibitemShut
  {NoStop}%
\bibitem [{\citenamefont {Flocken}\ \emph {et~al.}(1985)\citenamefont
  {Flocken}, \citenamefont {Guenther}, \citenamefont {Hardy},\ and\
  \citenamefont {Boyer}}]{flocken1985}%
  \BibitemOpen
  \bibfield  {author} {\bibinfo {author} {\bibfnamefont {J.~W.}\ \bibnamefont
  {Flocken}}, \bibinfo {author} {\bibfnamefont {R.~A.}\ \bibnamefont
  {Guenther}}, \bibinfo {author} {\bibfnamefont {J.~R.}\ \bibnamefont {Hardy}},
  \ and\ \bibinfo {author} {\bibfnamefont {L.~L.}\ \bibnamefont {Boyer}},\
  }\href {\doibase 10.1103/PhysRevB.31.7252} {\bibfield  {journal} {\bibinfo
  {journal} {Phys. Rev. B}\ }\textbf {\bibinfo {volume} {31}},\ \bibinfo
  {pages} {7252} (\bibinfo {year} {1985})}\BibitemShut {NoStop}%
\bibitem [{\citenamefont {Zhong}\ \emph {et~al.}(1995)\citenamefont {Zhong},
  \citenamefont {Vanderbilt},\ and\ \citenamefont {Rabe}}]{zhong1995}%
  \BibitemOpen
  \bibfield  {author} {\bibinfo {author} {\bibfnamefont {W.}~\bibnamefont
  {Zhong}}, \bibinfo {author} {\bibfnamefont {D.}~\bibnamefont {Vanderbilt}}, \
  and\ \bibinfo {author} {\bibfnamefont {K.~M.}\ \bibnamefont {Rabe}},\ }\href
  {\doibase 10.1103/PhysRevB.52.6301} {\bibfield  {journal} {\bibinfo
  {journal} {Phys. Rev. B}\ }\textbf {\bibinfo {volume} {52}},\ \bibinfo
  {pages} {6301} (\bibinfo {year} {1995})}\BibitemShut {NoStop}%
\bibitem [{\citenamefont {Vanderbilt}\ and\ \citenamefont
  {Zhong}(1998)}]{vanderbilt1998}%
  \BibitemOpen
  \bibfield  {author} {\bibinfo {author} {\bibfnamefont {D.}~\bibnamefont
  {Vanderbilt}}\ and\ \bibinfo {author} {\bibfnamefont {W.}~\bibnamefont
  {Zhong}},\ }\href {\doibase 10.1080/00150199808009158} {\bibfield  {journal}
  {\bibinfo  {journal} {Ferroelectrics}\ }\textbf {\bibinfo {volume} {206}},\
  \bibinfo {pages} {181} (\bibinfo {year} {1998})}\BibitemShut {NoStop}%
\bibitem [{\citenamefont {Benedek}\ and\ \citenamefont
  {Fennie}(2013)}]{Benedek2013}%
  \BibitemOpen
  \bibfield  {author} {\bibinfo {author} {\bibfnamefont {N.}~\bibnamefont
  {Benedek}}\ and\ \bibinfo {author} {\bibfnamefont {C.}~\bibnamefont
  {Fennie}},\ }\href {\doibase 10.1021/jp402046t} {\bibfield  {journal}
  {\bibinfo  {journal} {J. Phys. Chem. C}\ }\textbf {\bibinfo {volume} {117}},\
  \bibinfo {pages} {13339} (\bibinfo {year} {2013})}\BibitemShut {NoStop}%
\bibitem [{\citenamefont {Amisi}\ \emph {et~al.}(2012)\citenamefont {Amisi},
  \citenamefont {Bousquet}, \citenamefont {Katcho},\ and\ \citenamefont
  {Ghosez}}]{Amisi2012}%
  \BibitemOpen
  \bibfield  {author} {\bibinfo {author} {\bibfnamefont {S.}~\bibnamefont
  {Amisi}}, \bibinfo {author} {\bibfnamefont {E.}~\bibnamefont {Bousquet}},
  \bibinfo {author} {\bibfnamefont {K.}~\bibnamefont {Katcho}}, \ and\ \bibinfo
  {author} {\bibfnamefont {P.}~\bibnamefont {Ghosez}},\ }\href {\doibase
  10.1103/PhysRevB.85.064112} {\bibfield  {journal} {\bibinfo  {journal} {Phys.
  Rev. B}\ }\textbf {\bibinfo {volume} {85}},\ \bibinfo {pages} {064112}
  (\bibinfo {year} {2012})}\BibitemShut {NoStop}%
\bibitem [{\citenamefont {Kresse}\ and\ \citenamefont
  {Furthm{\"u}ller}(1996)}]{vasp1}%
  \BibitemOpen
  \bibfield  {author} {\bibinfo {author} {\bibfnamefont {G.}~\bibnamefont
  {Kresse}}\ and\ \bibinfo {author} {\bibfnamefont {J.}~\bibnamefont
  {Furthm{\"u}ller}},\ }\href {\doibase 10.1103/PhysRevB.54.11169} {\bibfield
  {journal} {\bibinfo  {journal} {Phys. Rev. B}\ }\textbf {\bibinfo {volume}
  {54}},\ \bibinfo {pages} {11169} (\bibinfo {year} {1996})}\BibitemShut
  {NoStop}%
\bibitem [{\citenamefont {Perdew}\ \emph {et~al.}(2008)\citenamefont {Perdew},
  \citenamefont {Ruzsinszky}, \citenamefont {Csonka}, \citenamefont {Vydrov},
  \citenamefont {Scuseria}, \citenamefont {Constantin}, \citenamefont {Zhou},\
  and\ \citenamefont {Burke}}]{Perdew2008}%
  \BibitemOpen
  \bibfield  {author} {\bibinfo {author} {\bibfnamefont {J.~P.}\ \bibnamefont
  {Perdew}}, \bibinfo {author} {\bibfnamefont {A.}~\bibnamefont {Ruzsinszky}},
  \bibinfo {author} {\bibfnamefont {G.~I.}\ \bibnamefont {Csonka}}, \bibinfo
  {author} {\bibfnamefont {O.~A.}\ \bibnamefont {Vydrov}}, \bibinfo {author}
  {\bibfnamefont {G.~E.}\ \bibnamefont {Scuseria}}, \bibinfo {author}
  {\bibfnamefont {L.~A.}\ \bibnamefont {Constantin}}, \bibinfo {author}
  {\bibfnamefont {X.}~\bibnamefont {Zhou}}, \ and\ \bibinfo {author}
  {\bibfnamefont {K.}~\bibnamefont {Burke}},\ }\href {\doibase
  10.1103/PhysRevLett.100.136406} {\bibfield  {journal} {\bibinfo  {journal}
  {Phys. Rev. Lett.}\ }\textbf {\bibinfo {volume} {100}},\ \bibinfo {eid}
  {136406} (\bibinfo {year} {2008})}\BibitemShut {NoStop}%
\bibitem [{\citenamefont {Gonze}\ and\ \citenamefont {Lee}(1997)}]{gonze1997}%
  \BibitemOpen
  \bibfield  {author} {\bibinfo {author} {\bibfnamefont {X.}~\bibnamefont
  {Gonze}}\ and\ \bibinfo {author} {\bibfnamefont {C.}~\bibnamefont {Lee}},\
  }\href {\doibase 10.1103/PhysRevB.55.10355} {\bibfield  {journal} {\bibinfo
  {journal} {Phys. Rev. B}\ }\textbf {\bibinfo {volume} {55}},\ \bibinfo
  {pages} {10355} (\bibinfo {year} {1997})}\BibitemShut {NoStop}%
\bibitem [{\citenamefont {Togo}\ \emph {et~al.}(2008)\citenamefont {Togo},
  \citenamefont {Oba},\ and\ \citenamefont {Tanaka}}]{phonopy}%
  \BibitemOpen
  \bibfield  {author} {\bibinfo {author} {\bibfnamefont {A.}~\bibnamefont
  {Togo}}, \bibinfo {author} {\bibfnamefont {F.}~\bibnamefont {Oba}}, \ and\
  \bibinfo {author} {\bibfnamefont {I.}~\bibnamefont {Tanaka}},\ }\href
  {\doibase 10.1103/PhysRevB.78.134106} {\bibfield  {journal} {\bibinfo
  {journal} {Phys. Rev. B}\ }\textbf {\bibinfo {volume} {78}},\ \bibinfo
  {pages} {134106} (\bibinfo {year} {2008})}\BibitemShut {NoStop}%
\bibitem [{\citenamefont {Gonze}\ \emph {et~al.}(2002)\citenamefont {Gonze},
  \citenamefont {Beuken}, \citenamefont {Caracas}, \citenamefont {Detraux},
  \citenamefont {Fuchs}, \citenamefont {Rignanese}, \citenamefont {Sindic},
  \citenamefont {Verstraete}, \citenamefont {Zerah}, \citenamefont {Jollet},
  \citenamefont {Torrent}, \citenamefont {Roy}, \citenamefont {Mikami},
  \citenamefont {Ghosez}, \citenamefont {Raty},\ and\ \citenamefont
  {Allan}}]{abinit}%
  \BibitemOpen
  \bibfield  {author} {\bibinfo {author} {\bibfnamefont {X.}~\bibnamefont
  {Gonze}}, \bibinfo {author} {\bibfnamefont {J.-M.}\ \bibnamefont {Beuken}},
  \bibinfo {author} {\bibfnamefont {R.}~\bibnamefont {Caracas}}, \bibinfo
  {author} {\bibfnamefont {F.}~\bibnamefont {Detraux}}, \bibinfo {author}
  {\bibfnamefont {M.}~\bibnamefont {Fuchs}}, \bibinfo {author} {\bibfnamefont
  {G.-M.}\ \bibnamefont {Rignanese}}, \bibinfo {author} {\bibfnamefont
  {L.}~\bibnamefont {Sindic}}, \bibinfo {author} {\bibfnamefont
  {M.}~\bibnamefont {Verstraete}}, \bibinfo {author} {\bibfnamefont
  {G.}~\bibnamefont {Zerah}}, \bibinfo {author} {\bibfnamefont
  {F.}~\bibnamefont {Jollet}}, \bibinfo {author} {\bibfnamefont
  {M.}~\bibnamefont {Torrent}}, \bibinfo {author} {\bibfnamefont
  {A.}~\bibnamefont {Roy}}, \bibinfo {author} {\bibfnamefont {M.}~\bibnamefont
  {Mikami}}, \bibinfo {author} {\bibfnamefont {P.}~\bibnamefont {Ghosez}},
  \bibinfo {author} {\bibfnamefont {J.-Y.}\ \bibnamefont {Raty}}, \ and\
  \bibinfo {author} {\bibfnamefont {D.}~\bibnamefont {Allan}},\ }\href
  {\doibase 10.1016/S0927-0256(02)00325-7} {\bibfield  {journal} {\bibinfo
  {journal} {Comp. Mater. Sci.}\ }\textbf {\bibinfo {volume} {25}},\ \bibinfo
  {pages} {478} (\bibinfo {year} {2002})}\BibitemShut {NoStop}%
\bibitem [{\citenamefont {Ghosez}\ \emph {et~al.}(1999)\citenamefont {Ghosez},
  \citenamefont {Cockayne}, \citenamefont {Waghmare},\ and\ \citenamefont
  {Rabe}}]{ghosez1999}%
  \BibitemOpen
  \bibfield  {author} {\bibinfo {author} {\bibfnamefont {P.}~\bibnamefont
  {Ghosez}}, \bibinfo {author} {\bibfnamefont {E.}~\bibnamefont {Cockayne}},
  \bibinfo {author} {\bibfnamefont {U.~V.}\ \bibnamefont {Waghmare}}, \ and\
  \bibinfo {author} {\bibfnamefont {K.~M.}\ \bibnamefont {Rabe}},\ }\href
  {\doibase 10.1103/PhysRevB.60.836} {\bibfield  {journal} {\bibinfo  {journal}
  {Phys. Rev. B}\ }\textbf {\bibinfo {volume} {60}},\ \bibinfo {pages} {836}
  (\bibinfo {year} {1999})}\BibitemShut {NoStop}%
\bibitem [{\citenamefont {Rabe}(2013)}]{KRabe2013a}%
  \BibitemOpen
  \bibfield  {author} {\bibinfo {author} {\bibfnamefont {K.~M.}\ \bibnamefont
  {Rabe}},\ }\enquote {\bibinfo {title} {Antiferroelectricity in oxides: A
  reexamination},}\ in\ \href {\doibase 10.1002/9783527654864.ch7} {\emph
  {\bibinfo {booktitle} {Functional Metal Oxides}}}\ (\bibinfo  {publisher}
  {{Wiley-VCH}},\ \bibinfo {year} {2013})\ pp.\ \bibinfo {pages}
  {221--244}\BibitemShut {NoStop}%
\bibitem [{\citenamefont {Glazer}(1972)}]{glazer1972}%
  \BibitemOpen
  \bibfield  {author} {\bibinfo {author} {\bibfnamefont {A.~M.}\ \bibnamefont
  {Glazer}},\ }\href {\doibase 10.1107/S0567740872007976} {\bibfield  {journal}
  {\bibinfo  {journal} {Acta Cryst. B}\ }\textbf {\bibinfo {volume} {28}},\
  \bibinfo {pages} {3384} (\bibinfo {year} {1972})}\BibitemShut {NoStop}%
\bibitem [{\citenamefont {Ghosez}\ \emph {et~al.}(1996)\citenamefont {Ghosez},
  \citenamefont {Gonze},\ and\ \citenamefont {Michenaud}}]{ghosez1996}%
  \BibitemOpen
  \bibfield  {author} {\bibinfo {author} {\bibfnamefont {P.}~\bibnamefont
  {Ghosez}}, \bibinfo {author} {\bibfnamefont {X.}~\bibnamefont {Gonze}}, \
  and\ \bibinfo {author} {\bibfnamefont {J.}~\bibnamefont {Michenaud}},\ }\href
  {http://iopscience.iop.org/0295-5075/33/9/713} {\bibfield  {journal}
  {\bibinfo  {journal} {Europhys. Lett.}\ }\textbf {\bibinfo {volume} {713}},\
  \bibinfo {pages} {713} (\bibinfo {year} {1996})}\BibitemShut {NoStop}%
\bibitem [{\citenamefont {Orobengoa}\ \emph {et~al.}(2009)\citenamefont
  {Orobengoa}, \citenamefont {Capillas}, \citenamefont {Aroyo},\ and\
  \citenamefont {Perez-Mato}}]{Orobengoa2009}%
  \BibitemOpen
  \bibfield  {author} {\bibinfo {author} {\bibfnamefont {D.}~\bibnamefont
  {Orobengoa}}, \bibinfo {author} {\bibfnamefont {C.}~\bibnamefont {Capillas}},
  \bibinfo {author} {\bibfnamefont {M.~I.}\ \bibnamefont {Aroyo}}, \ and\
  \bibinfo {author} {\bibfnamefont {J.~M.}\ \bibnamefont {Perez-Mato}},\ }\href
  {\doibase 10.1107/S0021889809028064} {\bibfield  {journal} {\bibinfo
  {journal} {J. Appl. Crys.}\ }\textbf {\bibinfo {volume} {42}},\ \bibinfo
  {pages} {820} (\bibinfo {year} {2009})}\BibitemShut {NoStop}%
\bibitem [{\citenamefont {Ghosez}\ \emph {et~al.}(1998)\citenamefont {Ghosez},
  \citenamefont {Michenaud},\ and\ \citenamefont {Gonze}}]{ghosez1998}%
  \BibitemOpen
  \bibfield  {author} {\bibinfo {author} {\bibfnamefont {P.}~\bibnamefont
  {Ghosez}}, \bibinfo {author} {\bibfnamefont {J.-P.}\ \bibnamefont
  {Michenaud}}, \ and\ \bibinfo {author} {\bibfnamefont {X.}~\bibnamefont
  {Gonze}},\ }\href {\doibase 10.1103/PhysRevB.58.6224} {\bibfield  {journal}
  {\bibinfo  {journal} {Phys. Rev. B}\ }\textbf {\bibinfo {volume} {58}},\
  \bibinfo {pages} {6224} (\bibinfo {year} {1998})}\BibitemShut {NoStop}%
\bibitem [{Note1()}]{Note1}%
  \BibitemOpen
  \bibinfo {note} {The eigendisplacements are normalized as follows: {$
  \delimiter "426830A \eta \delimiter 69640972 M \delimiter 86418188 \eta
  \delimiter "526930B = 1$}. $M$ is the atomic mass matrix $M_{ij} = \protect
  \sqrt {M_i M_j}$ where $M_i$ is the mass of atom $i$.}\BibitemShut {Stop}%
\bibitem [{Note2()}]{Note2}%
  \BibitemOpen
  \bibinfo {note} {The IFC matrix is defined by $C_{\alpha \beta i j}=\protect
  \frac {\partial F_{\alpha i}}{\partial r_{\beta j}}$ where $i$ and $j$ refers
  to the directions, $\alpha $ and $\beta $ to the atoms, $F$ is the force on
  an atom and $r$ the atomic position. The acoustic sum rule imposes $\DOTSB
  \sum@ \slimits@ _\beta C_{\alpha \beta i i}=0$. The on-site IFC of atom
  $\alpha $ along direction $i$ corresponding to the matrix element $C_{\alpha
  \alpha i i}$ is thus equal to $C_{\alpha \alpha i i}=-\DOTSB \sum@ \slimits@
  _{\beta \not =\alpha } C_{\alpha \beta i i}$ and so it can be seen as the
  resulting force on the atom $\kappa $ induced by all the other atoms when
  this atom $\kappa $ is displaced.}\BibitemShut {Stop}%
\bibitem [{\citenamefont {Kornev}\ and\ \citenamefont
  {Bellaiche}(2007)}]{kornev2007}%
  \BibitemOpen
  \bibfield  {author} {\bibinfo {author} {\bibfnamefont {I.~A.}\ \bibnamefont
  {Kornev}}\ and\ \bibinfo {author} {\bibfnamefont {L.}~\bibnamefont
  {Bellaiche}},\ }\href {\doibase 10.1080/01411590701228117} {\bibfield
  {journal} {\bibinfo  {journal} {Phase Trans.}\ }\textbf {\bibinfo {volume}
  {80}},\ \bibinfo {pages} {385} (\bibinfo {year} {2007})}\BibitemShut
  {NoStop}%
\bibitem [{\citenamefont {G\"unter}\ \emph {et~al.}(2012)\citenamefont
  {G\"unter}, \citenamefont {Bousquet}, \citenamefont {David}, \citenamefont
  {Boullay}, \citenamefont {Ghosez}, \citenamefont {Prellier},\ and\
  \citenamefont {Fiebig}}]{gunter2012}%
  \BibitemOpen
  \bibfield  {author} {\bibinfo {author} {\bibfnamefont {T.}~\bibnamefont
  {G\"unter}}, \bibinfo {author} {\bibfnamefont {E.}~\bibnamefont {Bousquet}},
  \bibinfo {author} {\bibfnamefont {A.}~\bibnamefont {David}}, \bibinfo
  {author} {\bibfnamefont {P.}~\bibnamefont {Boullay}}, \bibinfo {author}
  {\bibfnamefont {P.}~\bibnamefont {Ghosez}}, \bibinfo {author} {\bibfnamefont
  {W.}~\bibnamefont {Prellier}}, \ and\ \bibinfo {author} {\bibfnamefont
  {M.}~\bibnamefont {Fiebig}},\ }\href {\doibase 10.1103/PhysRevB.85.214120}
  {\bibfield  {journal} {\bibinfo  {journal} {Phys. Rev. B}\ }\textbf {\bibinfo
  {volume} {85}},\ \bibinfo {pages} {214120} (\bibinfo {year}
  {2012})}\BibitemShut {NoStop}%
\bibitem [{\citenamefont {Eklund}\ \emph {et~al.}(2009)\citenamefont {Eklund},
  \citenamefont {Fennie},\ and\ \citenamefont {Rabe}}]{Eklund2009}%
  \BibitemOpen
  \bibfield  {author} {\bibinfo {author} {\bibfnamefont {C.-J.}\ \bibnamefont
  {Eklund}}, \bibinfo {author} {\bibfnamefont {C.~J.}\ \bibnamefont {Fennie}},
  \ and\ \bibinfo {author} {\bibfnamefont {K.~M.}\ \bibnamefont {Rabe}},\
  }\href {\doibase 10.1103/PhysRevB.79.220101} {\bibfield  {journal} {\bibinfo
  {journal} {Phys. Rev. B}\ }\textbf {\bibinfo {volume} {79}},\ \bibinfo
  {pages} {220101} (\bibinfo {year} {2009})}\BibitemShut {NoStop}%
\end{thebibliography}%

\end{document}